\documentclass{PoS}

\usepackage{subfigure}

\title{Describing the Uncertainties in VLBI Images}

\ShortTitle{Describing the Uncertainties in VLBI Images}

\author{\speaker{Colm Coughlan}\\
        Department of Physics, University College Cork, Ireland\\
        E-mail: \email{colm.coughlan@umail.ucc.ie}}

\author{Denise Gabuzda\\
        Department of Physics, University College Cork, Ireland\\
        E-mail: \email{d.gabuzda@ucc.ie}}

\abstract{We are investigating a new approach to modelling uncertainties in individual pixels of Very Long Baseline Interferometry (VLBI) intensity (Stokes I, Q, U) images. Comparison of distributions of our calculated uncertainties for model sources with the results of Monte Carlo simulations shows that our method correctly reproduces the overall level and pattern of uncertainties in intensity images for model sources that are not too compact. Refinement of the approach to better reproduce uncertainties in compact sources is being studied. This new approach should ultimately provide a means to correctly estimate pixel-based uncertainties in the Stokes parameters I, Q and U, including the effect of correlations between values in different pixels.}

\FullConference{11th European VLBI Network Symposium \& Users Meeting,\\
		October 9-12, 2012\\
		Bordeaux, France}

\begin{document}

\section{Deconvolution with the CLEAN algorithm}

The CLEAN algorithm is one of the most popular deconvolution algorithms in radio astronomy. Originally developed by H\"{o}gbom 1974 \cite{hogbom}, there are many different versions of the algorithm in use today, including the Clark CLEAN algorithm \cite{clark} and multi-scale CLEAN algorithms \cite{multi}. Each of these algorithms has specific systematic errors associated with it. The question of how to reliably estimate uncertainties of the measured intensity at particular points of an overall intensity distribution is not trivial. The Clark CLEAN deconvolution algorithm essentially models the source as a sum of point sources ($\delta$ functions), which are convolved with the CLEAN beam to form the CLEAN map. This means that the value at each point formally depends on the values at all other points in the image containing significant flux. In this paper we attempt to describe the behaviour of the systematic errors in the standard Clark CLEAN algorithm for various model sources.

\section{A model for the errors}

In the final step of the Clark CLEAN algorithm, the CLEAN components are convolved with the restoring beam to create a partial CLEAN map. This can be represented mathematically as

\begin{equation}
\label{partialclean}
 I_{C_{k}} = \sum^{n}_{l=1}B(x_{k}-x_{l},y_{k}-y_{l})i_{l} \equiv \sum^{n}_{l=1}B_{lk}i_{l}
\end{equation}

\noindent
 where $i_{l}$ is the total flux of the CLEAN components at pixel $l$ and $B(x_{k}-x_{l},y_{k}-y_{l})$ is the value of the restoring beam at pixel $k$ (with coordinates $(x_{k},y_{k})$) when the beam is centred on pixel $l$ (with coordinates $(x_{l},y_{l})$). The residual map is then usually added to this map to give the final CLEAN map:

\begin{equation}
 I_{k} = I_{C_{k}} + R_{k}
\end{equation}

\noindent
It has been usual to adopt the root-mean-square (rms) deviations in the residual map (or in the final CLEAN map far from any regions containing real flux) $\sigma_{rms}$ as an estimate of the total uncertainty in the measured flux in an individual pixel. We instead consider this uncertainty to be comprised of a term associated with the CLEAN process, $\sigma_{CLEAN}$, and $\sigma_{rms}$ added in quadrature, similar to the treatment of Hovatta et al. \cite{hovatta}. Further, to determine $\sigma_{CLEAN}$, we hypothesized that the uncertainty in each of the ``merged'' CLEAN components (the sum of all CLEAN components at a given pixel) is approximately constant and given by

\begin{equation}
 \label{ccerror}
f \sigma_{rms}
\end{equation}

\noindent
where $f$ is a coefficient of order unity. This hypothesis does not have a rigorous mathematical basis in the sense of being derived from a full mathematical description of the entire CLEAN algorithm. In fact, the individual un-collapsed CLEAN components will be correlated with each other, and so will, in general, have correlated uncertainties. However, we will proceed on the hypothesis that the effect of the overall CLEAN process is to make the uncertainties in the ``merged'' CLEAN components approximately equal, and not significantly correlated. This allows the uncertainty in the flux $I$ at pixel $k$ of the final CLEAN map to be calculated using standard formulae for propagation of errors (see, for example, \cite{stats}):

\begin{equation}
 \label{prefinalerror}
 \sigma_{I_{C_{k}}}^{2} = f^{2} \sigma_{rms}^{2} \sum^{n}_{l=1} B^{2}_{lk}
\end{equation}

\begin{equation}
 \label{finalerror}
 \sigma_{I_{k}}^{2} = \sigma_{I_{C_{k}}}^{2} + \sigma_{rms}^{2}
\end{equation}

\section{Monte Carlo simulations}

\begin{figure}

  \begin{center}
  \subfigure[Cylindrical jet with an intrinsic linear decrease in intensity]{
    \includegraphics[scale=0.3]{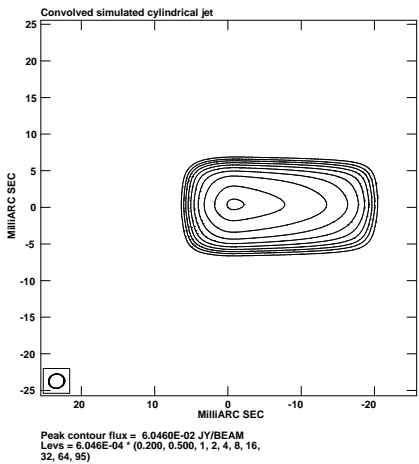}
    \label{cylindrical_imap}
  }
  \hfill
  \subfigure[Square-within-square source]{
    \includegraphics[scale=0.3]{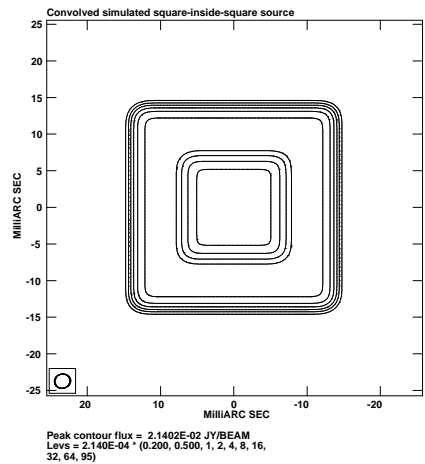}
    \label{square_imap}
  }
  \hfill
    \subfigure[Triple Gaussian source]{
    \includegraphics[scale=0.3]{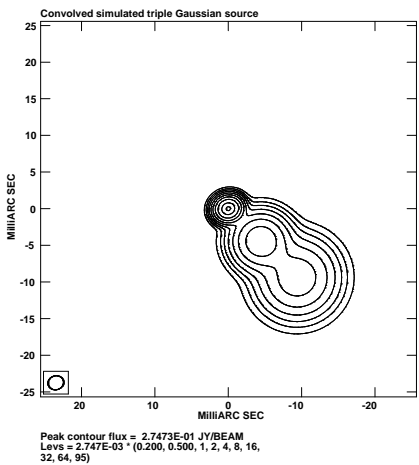}
    \label{tgc2_imap}
  }
  \hfill
    \subfigure[UV sampling fucntion]{
    \includegraphics[scale=0.3]{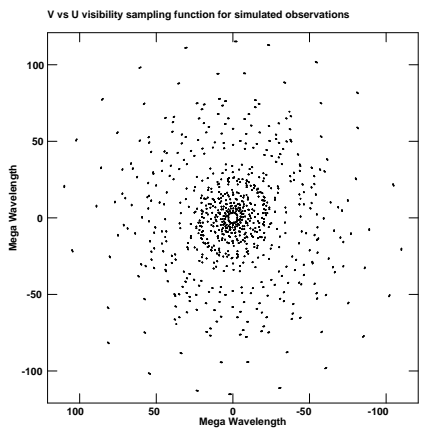}
    \label{uvcore}
  }
  \hfill
  
\caption{Convolved model maps and UV distribution used. Total flux in each case is 1 Jy. Beam = 2.14 x 1.93 mas, position angle $-75.11\,^{\circ}$. Contours shown are 0.2, 0.5, 1 , 2, 4, 8, 16, 32, 64, 95 \% of peak. Sources have peak values of (a) 60.46 mJy/Beam, (b) 21.40 mJy/Beam and (c) 274.7 mJy/Beam.}

  \end{center}
\end{figure}

Intensity distributions for several model sources were Fourier transformed and sampled using the baseline distribution from a typical real observation (Fig. \ref{uvcore}). Gaussian noise was added to the sampled visibilities. The ``noisy'' model visibility data were then imaged in the usual way in the AIPS package. One hundred such maps were made for each source model and compared to the original model after it had been convolved with an appropriate beam. This resulted in a Monte Carlo error map for each source model. Further, an uncertainty map was made for each of the 100 maps using Eqs. (\ref{prefinalerror})-(\ref{finalerror}). These 100 uncertainty maps were then averaged to create a calculated average error map which could be directly compared with the Monte Carlo error map.\\

\begin{figure}
  \subfigure[Monte Carlo error map]{
    \includegraphics[scale=0.3]{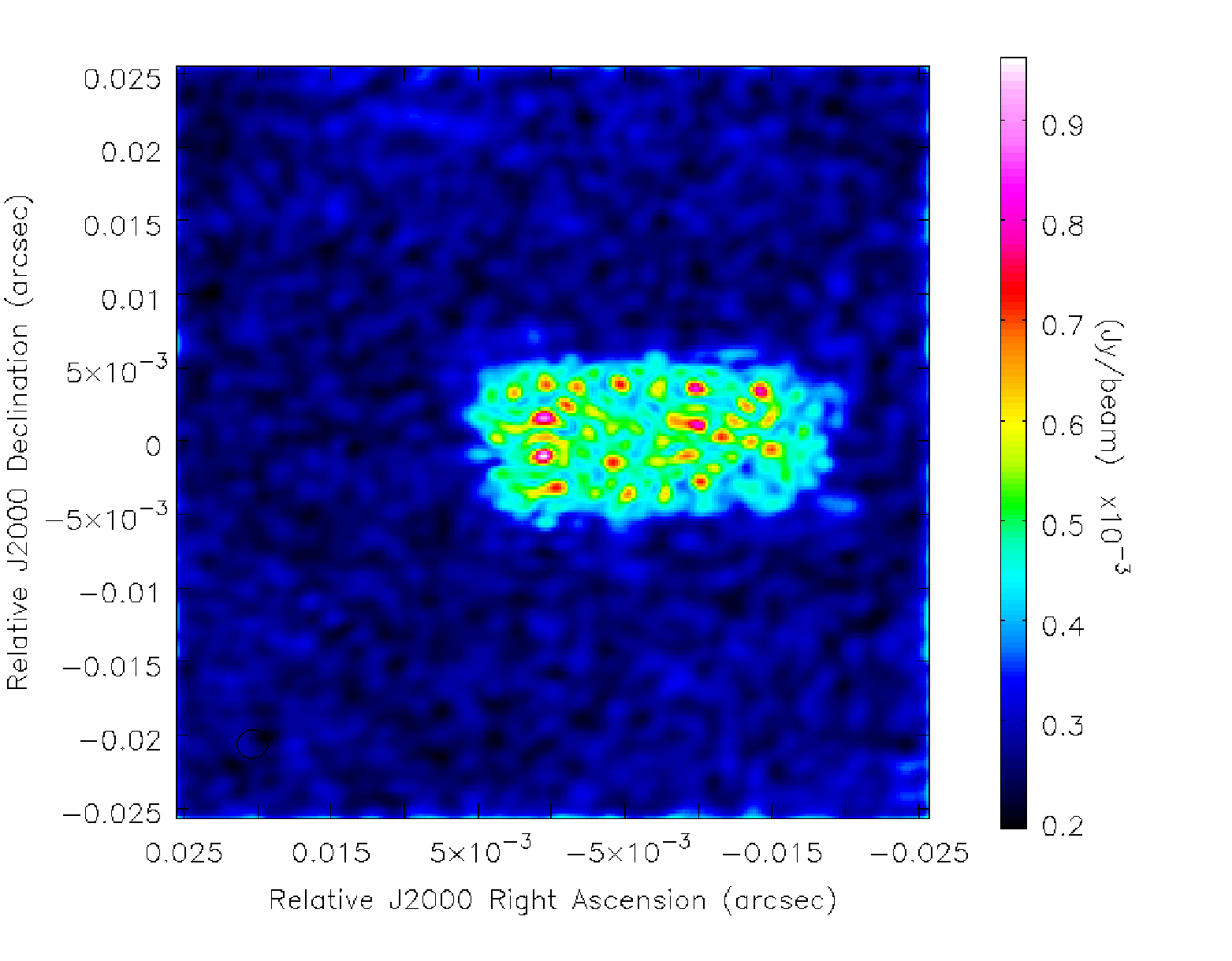}
  }
  \hfill
  \subfigure[Calculated error map]{
    \includegraphics[scale=0.3]{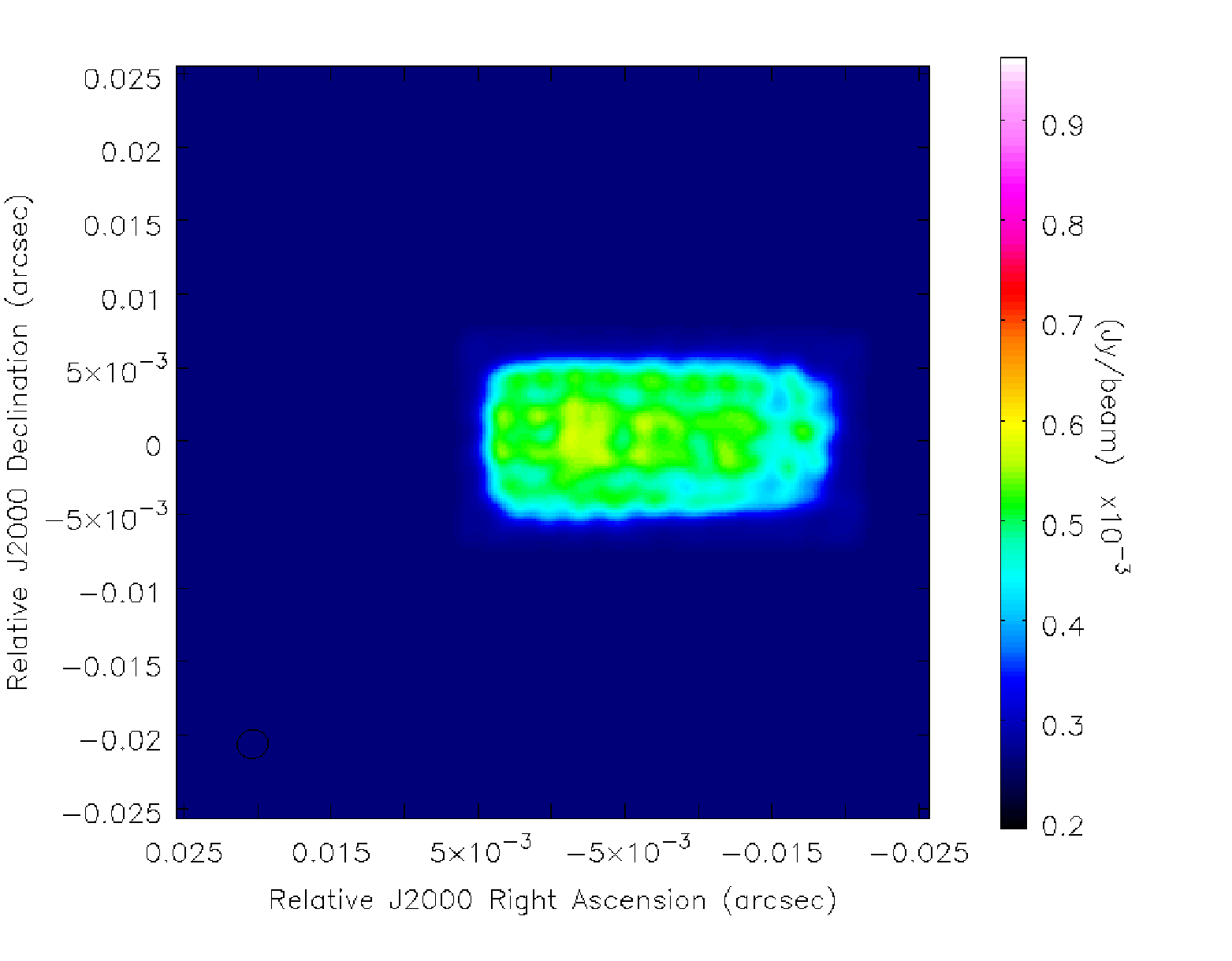}
  }
  \hfill
  
\caption{Monte Carlo and calculated error maps for a simulated cylindrical jet with a linear decrease in intensity.}
  \label{eoinjet}
\end{figure}

\begin{figure}
\begin{center}

  \end{center}
  \subfigure[Monte Carlo error map]{
    \includegraphics[scale=0.3]{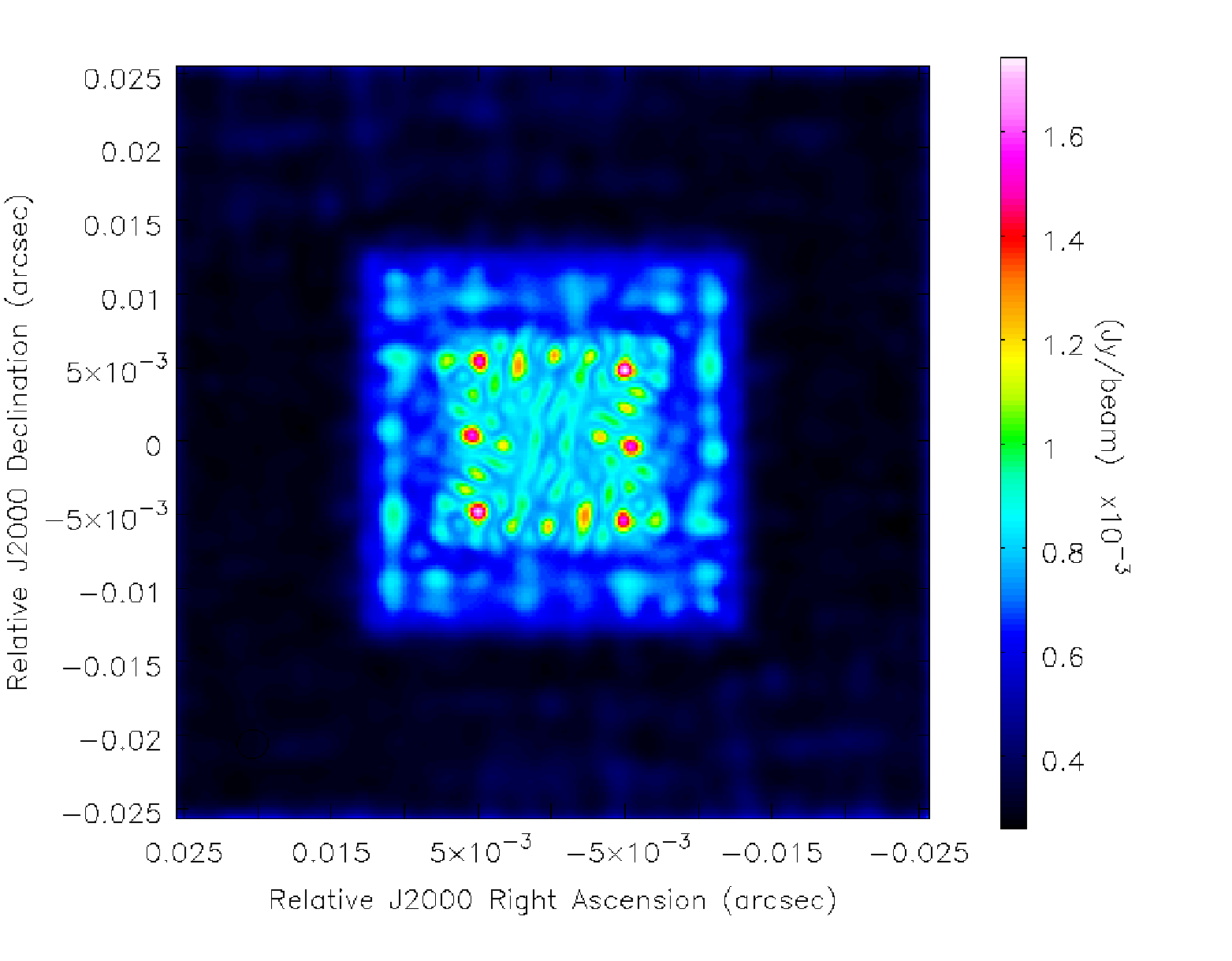}
  }
  \hfill
  \subfigure[Calculated error map]{
    \includegraphics[scale=0.3]{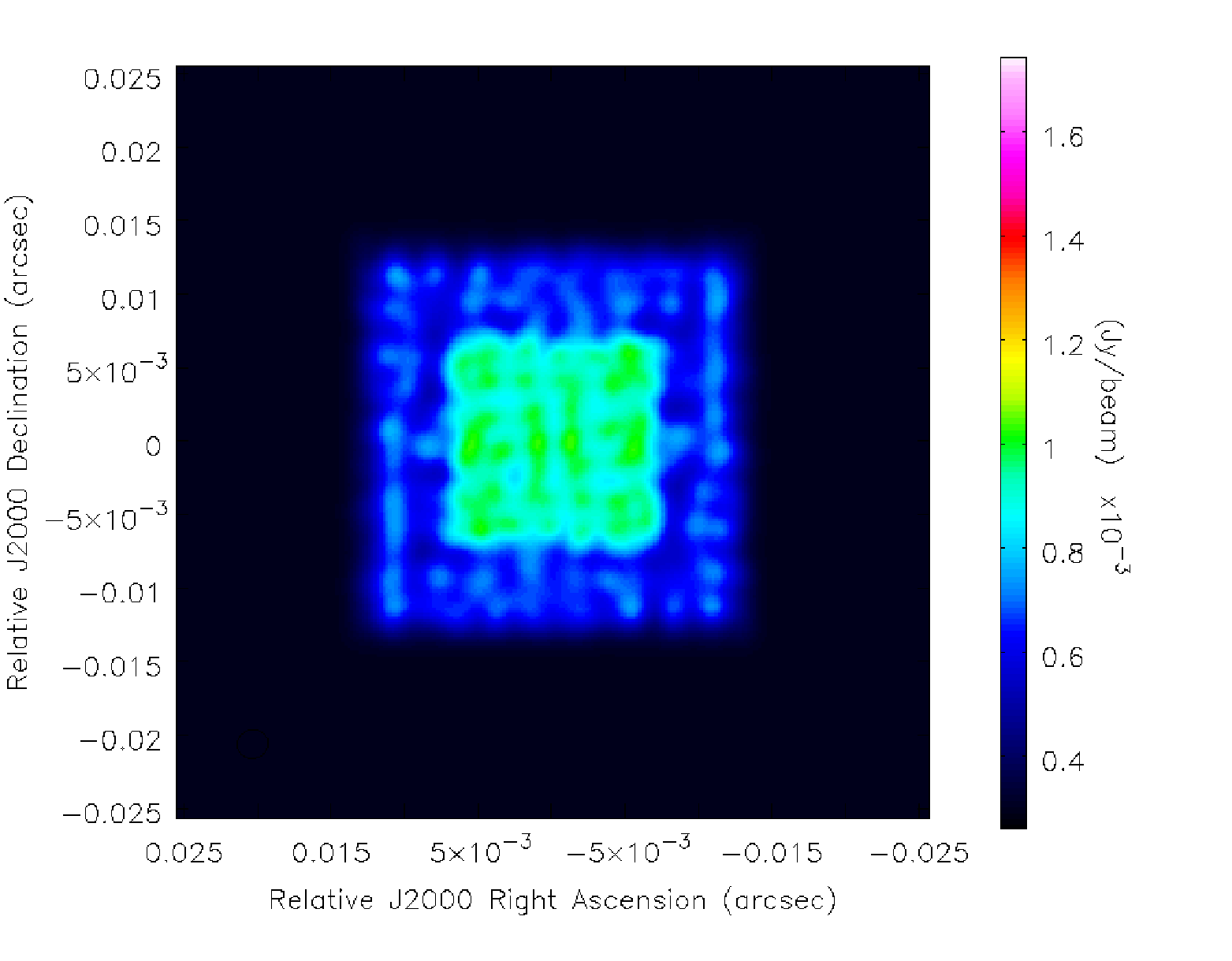}
  }
  \hfill
  
\caption{Simulated square-inside-square source.  This is not a realistic source, but highlights how the CLEAN algorithm responds to flat regions of different flux.}
  \label{sqjet}

\end{figure}

\begin{figure}
  \subfigure[Monte Carlo error map]{
    \includegraphics[scale=0.3]{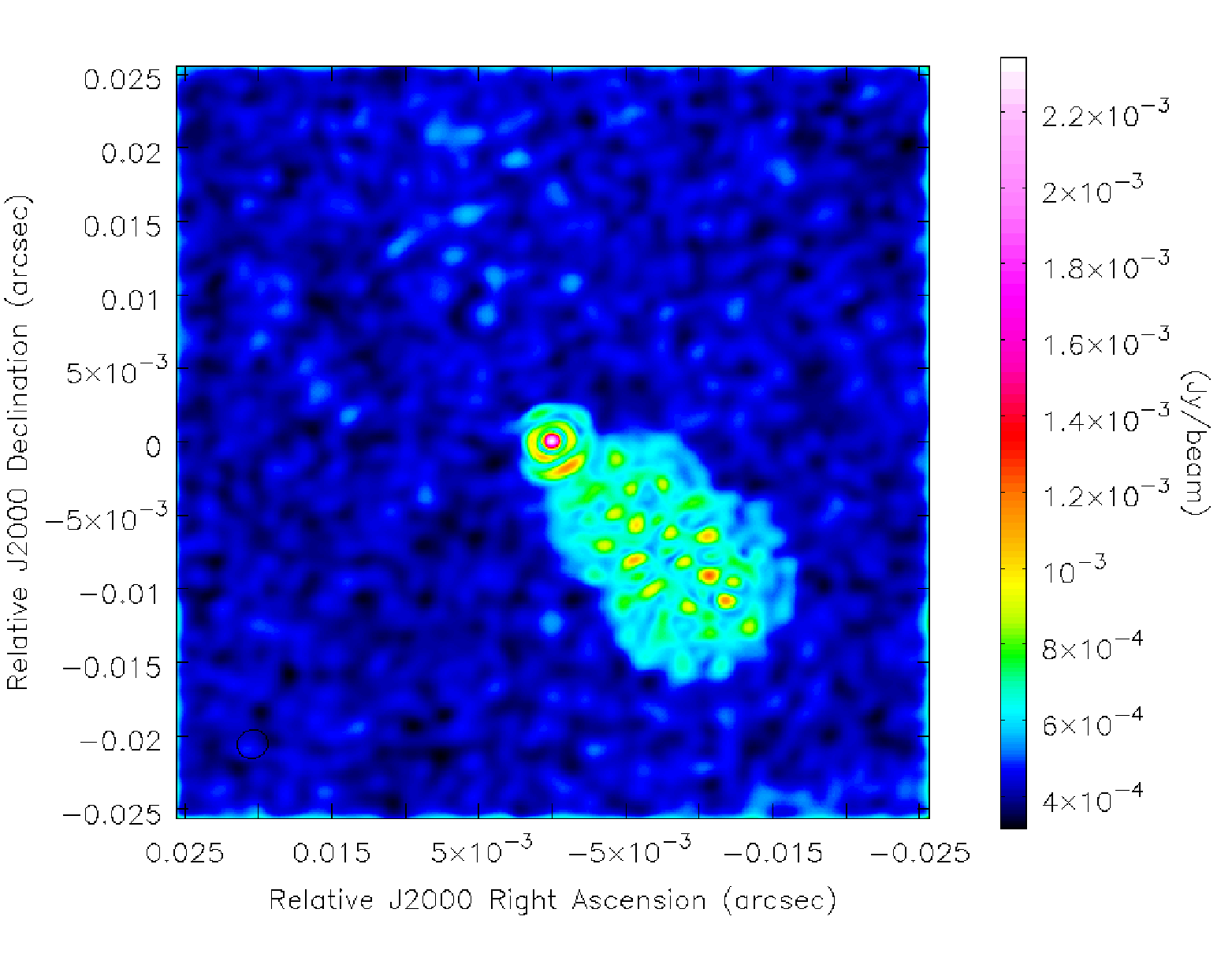}
  }
  \hfill
  \subfigure[Calculated error map]{
    \includegraphics[scale=0.3]{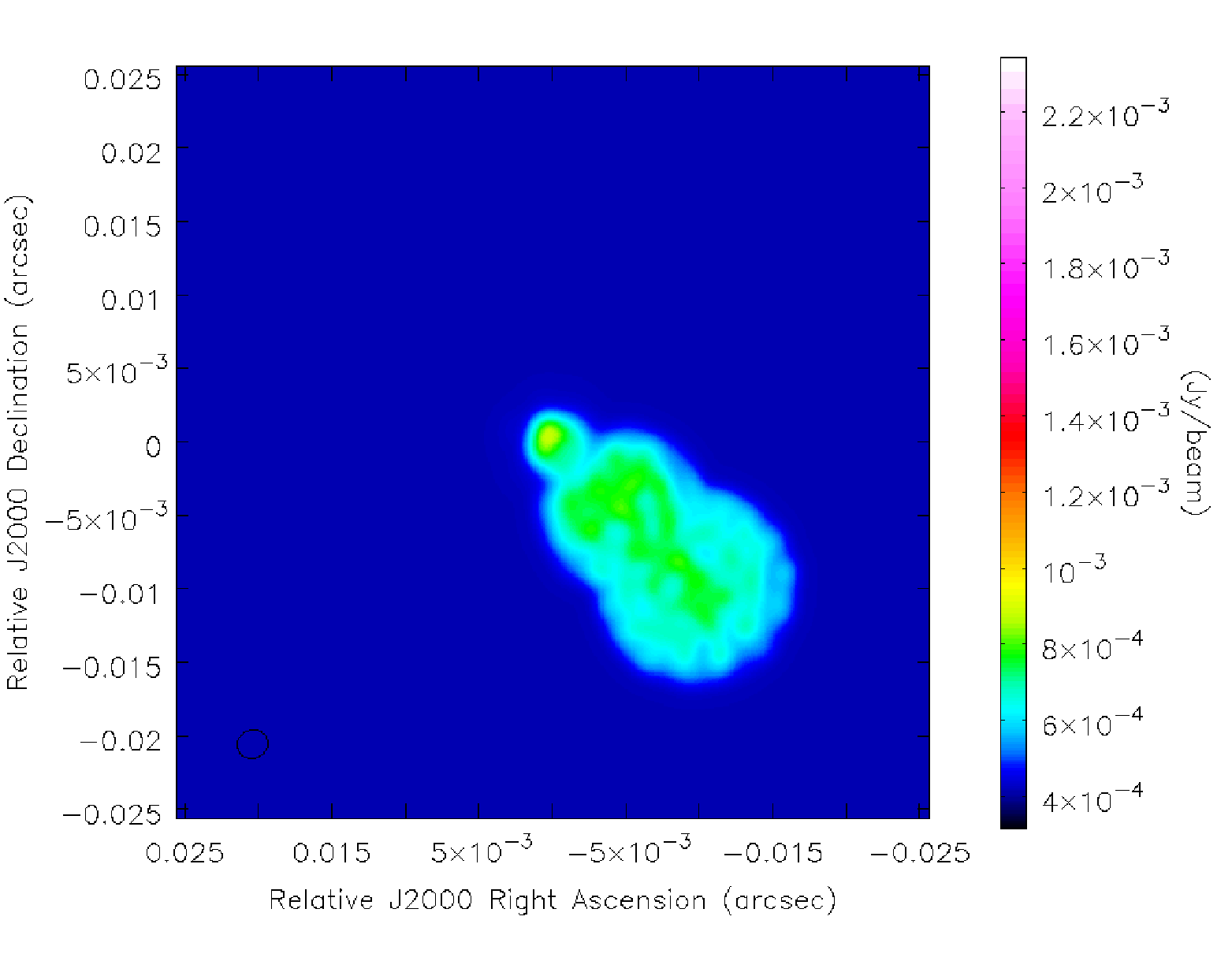}
  }
  \hfill
  
\caption{Monte Carlo and calculated error maps for a simulated core-jet source comprised of 3 Gaussian components.}
  \label{tgc2jet}

\end{figure}

\noindent
Three models are considered: a cylindrical jet-like source (Fig. \ref{cylindrical_imap}), a square region of constant intensity with another square region of higher constant intensity at its centre (Fig. \ref{square_imap}) and a triple-Gaussian core-jet-like source (Fig. \ref{tgc2_imap}). The cylindrical and square sources are not intended to represent real emission, but to highlight the systematic response of the CLEAN algorithm to certain types of structure. In all three cases,  comparisons between the Monte Carlo and calculated error maps show that the patterns displayed agree well, sometimes even in details (Figs. \ref{eoinjet} to \ref{tgc2jet}). This demonstrates that the uncertainty patterns are determined to a considerable extent by the distribution of CLEAN components. Good quantitative agreement between the calculated and Monte Carlo uncertainties is achieved with $f=0.5$.\\

\noindent
In particular the square-within-square source (Fig. \ref{sqjet}) highlights the fact that regions of considerably different flux can have somewhat different uncertainties, and the shape of the region and possibly the presence of sharp edges also has an effect on the uncertainty (note the peaks in the Monte Carlo error map at the points of the inner square).\\

\noindent
It is important to note however that these sources are all fairly extended. Such good agreement between the Monte Carlo and calculated error maps is not seen for more compact sources, such as ``delta function'' sources and very small Gaussians. The CLEAN algorithm can be expected to be better suited to these types of sources, as its modelling of the sources as a series of delta functions becomes a better approximation. Eqs. (\ref{prefinalerror})-(\ref{finalerror}) predict uncertainty distributions that are too smooth and extended in these cases. Further work is ongoing on generalising the equations in Section 2 to be able to better reproduce uncertainties in images of more compact sources.

\section{Conclusion}

Our error model provides good estimates of the uncertainties in individual pixels in Stokes I, Q and U images of sources that are not too compact, as well as a quantitative approach to describing the relationships (correlations) between values in different pixels. In our model these correlations come about because each pixel value is calculated using all the CLEAN components after each has been convolved with the CLEAN beam. Knowledge of the uncertainty in each pixel, as well as the correlations between nearby pixels, should allow the rigorous calculation of various further uncertainties, taking into account relevant correlations between pixels, such as image values averaged over some number of neighbouring pixels and various derived quantities, such as spectral index, polarised flux, degree of polarisation, polarisation angle and Faraday rotation measure. \\

This enables a wide range of analyses that have been hindered previously by inability to reliably compare values at different locations in images in the absence of reliable error estimates, in particular, analyses of gradients of the spectral index, degree of polarization and Faraday rotation. However the fact that the model does not currently work well for the compact structures observed in many VLBI sources means that further work will have to be done before it can be successfully applied to VLBI observations of compact AGN.

\section*{Acknowledgements}

\noindent
This work was supported by the Irish Research Council for Science, Engineering and Technology (IRCSET).

\end{document}